\documentclass{article}

\usepackage{arxiv}
\usepackage{amsmath,amssymb}
\usepackage[utf8]{inputenc} % allow utf-8 input
\usepackage[T1]{fontenc}    % use 8-bit T1 fonts
\usepackage{hyperref}       % hyperlinks
\usepackage{url}            % simple URL typesetting
\usepackage{booktabs}       % professional-quality tables
\usepackage{amsfonts}       % blackboard math symbols
\usepackage{nicefrac}       % compact symbols for 1/2, etc.
\usepackage{microtype}      % microtypography
\usepackage{lipsum}
\usepackage{graphicx}
\title{Computational Diplomacy: How ``Hackathons for Good'' Feed a Participatory Future for Multilateralism in the Digital Age}

\author{
 Thomas Maillart \\
  Geneva School of Economics and Management\\
  University of Geneva\\
  Geneva, Switzerland\\
  \texttt{thomas.maillart@unige.ch}\\
 \And
 Lucia Gomez \\
  Geneva School of Economics and Management\\
  University of Geneva\\
  Geneva, Switzerland\\
 \And
 Ewa Lombard \\
  MBS Business School\\
  Montpellier, France\\
 \And
 Alexander Nolte \\
  Eindhoven University of Technology\\
  Eindhoven, The Netherlands\\
  Carnegie Mellon University\\
  Pittsburgh, PA, USA\\
 \And
 Francesco Pisano \\
  United Nations Library and Archives of Geneva\\
  Geneva, Switzerland\\
}

\begin{document}
\maketitle
\begin{abstract}
This article explores the role of {\it hackathons for good} in building a community of software and hardware developers focused on addressing global SDG challenges. We theorise this movement as computational diplomacy: a decentralised, participatory process for digital governance that leverages collective intelligence to tackle major global issues. Analysing Devpost and GitHub data reveals that 30\% of hackathons since 2010 have addressed SDG topics, employing diverse technologies to create innovative solutions. Hackathons serve as crucial kairos moments, sparking innovation bursts that drive both immediate project outcomes and long-term production. We propose that these events harness the neurobiological basis of human cooperation and empathy, fostering a collective sense of purpose and reducing interpersonal prejudice. This bottom-up approach to digital governance integrates software development, human collective intelligence, and collective action, creating a dynamic model for transformative change. By leveraging kairos moments, computational diplomacy promotes a more inclusive and effective model for digital multilateral governance of the future.

\end{abstract}

\section{Introduction}
\label{sec:intro}
In recent years, a staggering number of environmental \cite{richardson_earth_2023}, technological \cite{wheatley_extreme_2016,kwon_three_2023,leveringhaus_whats_2018}, socio-economic \cite{svolik_polarization_2019} and geopolitical crises \cite{aviv_russian-ukraine_2023}, somehow linked together, have unravelled. These crises have affected millions of people, destabilised states, and heavily challenged multilateralism \cite{hosli_future_2021}. The multilateral framework was developed to foster international cooperation and prevent conflicts, recognising that many world issues transcend national borders and necessitate collective action beyond bi-lateral diplomatic relationships between nation states.\footnote{The multilateral system was developed in the nineteenth century through the Congress of Vienna (Nov. 1814 to June 1815), the Conference of Berlin (1884) and with the establishment of the Permanent Court of Arbitration (now the International Court of Justice) in the Hague in 1899} The establishment of the League of Nations and of the United Nations (UN), after respectively World Wars I and II marked a significant advancement, setting a framework for global diplomacy and international law, guided by principles such as sovereign equality and peaceful dispute resolution \cite{hosli_future_2021}. Recognising the urgency of reforming itself in the face of the aforementioned challenges, the UN have called to hold the Summit of the Future, deemed as a one-of-a-kind moment to redefine and strengthen multilateralism for a better future (c.f., Supplementary Materials). At its core, the Summit of the Future is driven by the vision laid out in the Secretary-General's 2021 report, ``Our Common Agenda,'' which advocates for an inclusive, networked, and efficient form of multilateralism. The resulting Pact for the Future is seen as a commitment among nations to leverage global tools to address emergent issues proactively. One of the six needed shifts, called {\it Rebuild Trust in Multilateralism}, emphasised during the preparatory discussions for the Summit\footnote{\url{https://highleveladvisoryboard.org/}} involves new forms of people engagement and collective intelligence, crucial for rebuilding trust in multilateralism. This shift acknowledges the imperative of involving a broader spectrum of society in shaping and decision-making processes, particularly the youth. With over half of the world's population under 30, the need for young people to have a meaningful seat at the decision-making table is highlighted. This approach is not just about consulting young people but involves recognising them as true and equal partners in crafting a sustainable future.

Historically, the multilateral system has
been shaped by the complex power-brokering dynamics of nation states aiming to serve their own interests \cite{martin_interests_1992}, which are influenced by a range of factors that may not always align with the interests of their respective population or people in other states. The rise of non-state actors and the increasing role of private economic actors in the global political economy further complicates this dynamics\cite{ kuttner_managed_1990} and the changing nature of international security, with a shift towards regional dynamics, adds another layer of complexity \cite{alagappa_regionalism_1995}. Nevertheless, in order to build resilience in the face of crises and transition challenges \cite{novalia_theorising_2020}, international organisations (IOs) need to engage with citizens in radically new ways \cite{woods_good_1999}. Participatory processes have attracted much attention as they allow at once to design {\it bottom-up} solutions and build social consensus around these solutions. They refer to methods and practices that actively involve stakeholders, especially those who are directly affected, in decision-making or problem-solving activities. This approach emphasises the importance of including a diverse range of voices and perspectives to ensure that outcomes are equitable and reflective of the needs and interests of all participants. Participatory processes are used in various fields, including community planning, policy-making, and organisational development, with the goal of enhancing transparency, accountability, and democracy \cite{taylor2018everybody,hou2017hacking}. They foster a sense of ownership and empowerment among participants, leading to more sustainable and accepted solutions. 

In this work, we consider how {\it hackathons for good}, a special form of public participation, inherited from, and having fundamentally shaped the digital revolution, can be particularly useful not only to advance concretely the agenda of the Summit of the Future, in particular Rebuilding Trust in Multilateralism, but also to fundamentally change perspectives on how the humankind shall envision how to \textit{embrace} its existential challenges instead of \textit{tackling} them.

To advance empirically on this question, it is key to recognise that contemporary participatory processes are more a cause of the digitisation of society, rather than a consequence. Indeed, participatory processes are at the root of the ``hacker'' movement, which was born in the late 1950s and has profoundly shaped the digital revolution \cite{levy_hackers_2010,coleman_coding_2012}. Historically, the hacker movement takes its deep roots back to the emergence of a strong post World War II counterculture born out of a disillusion of American intellectuals following the atomic bombings of Hiroshima and Nagasaki, and more broadly by a rejection of war \cite{turner_democratic_2015}. As information technologies developed, this counterculture morphed into the cyberculture \cite{turner_counterculture_2008} and nowadays into participation \cite{benkler_penguin_2011} as a \textit{condition} of the digital age with its promises and challenges \cite{barney_participatory_2016}.

As multilateral diplomacy primarily focuses on discussing and building international norms and agreements that promote global cooperation, peace, and address transnational challenges collectively, envisioning the future of multilateralism requires to contemplate that in the digital age \textit{code is law} \cite{lessig_code_2006}. While international forums, treaties, laws, policies and their enforcement certainly still have an important role to play in regulating the world's affairs, multilateral institutions shall reckon that software code and hardware designs deserve high attention in the way they are produced, as they enable and constrain people, nation states, regional blocs and international organisations alike, on their quest to overcome big challenges for the humankind. We call \textit{Computational Diplomacy} this process of developing source code and hardware designs with impact on the multilateral agenda. Although we recognise that other private entities and public organisations definitely use software and hardware to promote their computational diplomacy agenda, here, we investigate how participatory open source processes can support code development for good, involving a largely decentralised community, collectively building capacity to mass produce software pieces aimed at contributing to addressing some of the most pressing issues identified by the multilateral system, commonly known as the UN Sustainable Development Goals (SDGs). 

We recognise that other researchers have recently considered their own way of characterising computational diplomacy, from representing the complex bodies and actions in the multilateral system using multi-layered networks \cite{wernli_fostering_2023}, to computational and data science methods for conflict research \cite{cederman_computational_2023,chadefaux_automated_2023,galam_dynamics_2023,holyst_why_2023}, to using data science to support international negotiations \cite{cafiero_datafying_2023}, to robust automated global challenge response \cite{groen_facilitating_2023}, to designing democracy by design \cite{helbing_democracy_2023}. In this work, we are largely inspired by the latter view on computational diplomacy. However, we don't advocate for a new model. We rather argue and bring evidence that computational diplomacy has long started as an ecosystemic participatory process involving hundreds of thousands, if not millions, of citizens, who have focused on developing software code for advancing the SDGs. 

At the heart of our considerations for a social proxy of computational diplomacy and thus multilateralism, are hackathons. Hackathons --- which is a portmanteau of “hacking” and “marathon” --- are peculiar yet widespread events, where individuals form teams and collaborate on projects that are of interest to them \cite{falk2022future} for a short and intense ``kairos'' period of time \cite{orlikowski_its_2002}. These projects often aim to create solutions to concrete problems \cite{briscoe2014digital} and promise to give rise to long-term collaborative efforts and networks. Starting as collaborative coding events, hackathons have over time become increasingly popular also in non-technical fields offering participation opportunities for individuals to put their sense purpose into collective action~\cite{taylor2018everybody}. As they involve production and innovation through collective intelligence, they somehow go beyond what is commonly understood as public participation. We theorise that hackathons are a uniquely powerful type of temporary collaboration \cite{trainer2016hackathon,mendes2022socio}, particularly suited to building a long-term actionable form of ecosystemic participatory processes, which map into a broader range of social dynamics of cooperation and can be understood and modelled from a social physics perspective \cite{jusup_social_2022}.

To make an initial characterisation of computational diplomacy, we analysed an aggregated dataset that originated from the hackathon database Devpost\footnote{\href{https://devpost.com/}{https://devpost.com/}} together with event data from Github.\footnote{\href{https://github.com/}{https://github.com/}} The dataset consisted of $5,456$ distinct hackathons, $184,652$ unique projects, $290,795$ individual participants and $3,382,144$ software development events (from 7,643 projects), out of which $1,320$ hackathons (respectively, $37,647$ projects and $139,569$ participants) were associated with at least one of the seventeen SDGs. We found that the studied corpus of hackathons, indeed largely refers to the SDGs, with insights on how technology is used across SDGs. Assessing participation and the activity generated by hackathons, we find that they play a significant role, not only in enlarging and reinforcing communities, but also in fostering software and hardware code development.

The remainder of the article is organised as follows. In Section \ref{sec:background}, we review the advancements in participatory processes, quantitative research on open source communities and hackathons, and open on the neurobiological basis for purpose and collective innovation. In Section \ref{sec:theory}, we outline the theory of computational diplomacy, and formulate hypotheses to be tested. In Section \ref{sec:results}, we show how the concept of computational diplomacy unravels in a large heterogeneous community has managed to structure around participation to hackathons organised for advancing the SDGs. In Section \ref{sec:discussion}, we (i) discuss our results including (ii) their limitations. We also (iii) open a perspective on the possible psycho-biological fundamentals of hackathons as triggers for collective action for purpose and (iv) we outline three concrete cases for the future of computational diplomacy. In Section \ref{sec:conclusion}, we conclude and emphasise the importance of considering the future of multilateral diplomacy in a digitised world from the perspective of a highly decentralised, eco-systemic and participatory process as an additional form of (digital) governance for the world.

\section{Background}
\label{sec:background}

Our contribution on participatory processes for digital multilateral governance is built at the interface of two phenomena: (i) {\it peer production} and (ii) {\it kairos participatory moments}, which depending on the circumstances may be totally separated or, on the contrary, may operate mutually. In the case of computational diplomacy, we are indeed interested in the latter case. We also consider (iii) the possible neurobiological bases underlying kairos moments.

\subsection{Peer production, collective action and productive bursts}

The idea of peer-production takes its roots in attempts by scholars in the late 1990s and early 2000s to bring more understanding to a phenomenon known as the open source software (OSS) movement, which had become a new incarnation of the hacker movement with the invention of the open source GNU General Public Licence (GPL) by Richard Stallman in 1989 \cite{levy_hackers_2010}. The first prominent account of the open source movement was made by Eric Raymond in 1999. In \cite{raymond_cathedral_1999}, Raymond performed a comparative analysis between traditional top-down work organisation (i.e., ``The Cathedral'') and the open source bottom-up way (i.e., ``The Bazaar''), arguing that the bazaar model, characterised by decentralised, collaborative, and transparent development, would allow for rapid bug fixing, flexibility, and innovation through diverse contributions. It would also foster creativity and user engagement, closely aligning software with user needs by leveraging the collective intelligence of a broad developer community. In the {\it Coase's Penguin, or Linux and ``The Nature of the Firm'' }\cite{benkler_coases_2002}, first described {\it task self-selection} and {\it peer-review} as the fundamental ingredients of what he would call {\it peer-production}, arguing that a novel form of production organisation was born, different from {\it market} and {\it firms}, the two traditional models of production as devised by Ronald Coase in {\it The Nature of the Firm} \cite{coase_nature_1937}. In 2011, Benkler showed how peer-production, as a by-product of digitisation, had the potential to alter society for the best \cite{benkler_penguin_2011}. The open source movement has also attracted the interest of economists and management science scholars, primarily on the controversial questions of extrinsic versus intrinsic motivations to contribute \cite{tirole_simple_2002} and social practices \cite{von_krogh_carrots_2012}, as well as a vivid example of commons \cite{Ostrom2014} built at scale \cite{osterloh_open_2007}. At the organisational level, one aspect is key regarding public participation: open source principles have democratised development processes by lowering barriers to entry and enabling more inclusive groups of participants. This inclusion has fostered greater innovation and accelerated development cycles \cite{bosu_process_2017}. 

The success of the open source movement has also been documented with an effort to characterise and quantify the ``bazaar'' described by Raymond \cite{raymond_cathedral_1999}. One aspect is modularity: investigating the modular network of Debian Linux dependency ecosystem, Maillart et al. \cite{maillart_empirical_2008} found that the statistical mechanisms of {\it in-degree} use of software packages follows multiplicative stochastic process of proportional growth (a.k.a. preferential attachment), with continuous arrival and disappearance of packages. The capacity to produce vast quantities of software, furthermore in a modular way hence allowing for the quick low cost replacement of components, exhibits the efficiency of open source ecosystems with hundreds of thousands, if not millions, of software components that could potentially be dynamically re-combined in a creative destruction process \cite{maillart_empirical_2008}. 

Another consideration of interest to assert the economic efficiency of open source is the productivity of communities. This topic has remained highly controversial because it has been difficult to directly measure the effort behind each contribution. By studying bug bounty programs, a special form of peer-production leaning close to crowd-sourcing for cybersecurity and for which it is possible to measure productivity against incentives, it was shown that individual productivity decreases as people concentrate on searching bugs in one software, even if they have been successful at finding some in the past in that same software \cite{maillart_given_2017}. The authors found that indeed {\it ``given enough eyeballs, all bugs are shallow''} as proposed by Raymond in his seminal paper \cite{raymond_cathedral_1999}, but with some subtle limitations in the extent of individual contributions, possibly including cognitive load \cite{sweller_cognitive_1988}: It was advocated that contributors should rotate to generate a diversity of perspectives for performance \cite{hong_groups_2004}, which happens to bring support and economic justification for peer-production \cite{benkler_coases_2002}. Recently, Gillard et al. investigated MISP, an open source cybersecurity platform \cite{gillard_efficient_2022}. They found that larger crowds in participatory processes and modular organisation increase performance (in that case response time to cybersecurity threats was the independent variable). Performance can also be measured in terms of work quality. Using bi-partite recursive networks, it was shown how Wikipedia thematic sub-communities would create value together by revealing the specific structures of cooperation between contributors \cite{klein_virtuous_2015}. Reminiscent of \cite{scholtes_aristotle_2016}, they found that more contributors in a sub-community may also create dis-value.

Sornette et al. \cite{sornette_how_2014} tackled the productivity problem from statistical mechanics viewpoint as they found highly non-linear contribution dynamics with a scaling law of positive returns of scale,

\begin{equation}
  R \sim c^{\beta}~~with~~\beta \approx 4/3,
  \label{eq:scaling_productivity}
\end{equation}

with $R$ the number of commits over a short period of 5 days and $c$ the number contributors over the same period. Rationalising this surprising result, which by the way was found to be the first empirical rationalisation of the famous Aristotle's adage {\it the whole is more than the sum of its parts}, it was found that critical cascades of individual and collective contributions generate these productive bursts. Although Scholtes et al. \cite{scholtes_aristotle_2016} contended the results arguing for a {\it Ringelmann} effect of decreasing returns of scale, and a debate ensued with discussions on selection biases in data \cite{maillart_aristotle_2019}. Muri\'c et al. settled the issue independently in favour of productive bursts and positive returns of scale \cite{muric_collaboration_2019}, consecrating the paramount importance of kairos moments \cite{orlikowski_its_2002} in peer-production. In a nutshell, one can consider that productivity gains of peer-production are largely the result of kairos moments, and {\it vice-versa} : average productivity is high as a result of large deviation, highly non-linear burst of activity. 

Sornette et al. envisioned that productive bursts in open source communities are spontaneous \cite{sornette_how_2014,maillart_aristotle_2019}. However, one may also consider that some sorts of exogenous shocks \cite{crane_robust_2008} may trigger these bursts. While open source communities routinely interact online, they also periodically meet at (un)conferences and hackathons on physical premises, as well as online. Even though evidence remains to be established, these short kairos moments of a few hours to a few days where serendipity, diversity \cite{hong_groups_2004} and social interactions produce collective intelligence \cite{woolley_evidence_2010} may intuitively appear as good candidates for productive bursts \cite{sornette_how_2014}. We particularly focus on hackathons, because they are meant to produce, rather than showcasing (conferences) or discussing (negociations).

\subsection{Hackathons as Collaborative Kairos Moments}

The term {\it hackathon} was originally coined around the year 2000. At the time, it mainly referred to competitive coding events during which (often young) developers formed ad-hoc teams to work on a common project and compete for a small prize with a chance to land a future job \cite{briscoe2014digital}. Hackathons have proliferated into various domains including entrepreneurship \cite{angarita2023startup}, corporations \cite{pe2022corporate}, science \cite{huppenkothen2018hack}, (higher) education \cite{gama2018hackathon} and others. This adoption has also contributed to broadening the focus of hackathons from the development of (innovative) technology \cite{briscoe2014digital} to fostering learning \cite{schulten2024we} and community building \cite{nolte2020support}. Hackathons that focus on civic and environmental issues have received growing attention \cite{hou2017hacking}. These events --- which are often organised under the umbrella of ``hack for good’’ --- can serve as a means for the general public to contribute to solving issues they care about \cite{Lombard2024}. While events often take a technological solutionism perspective with its load of controversy \cite{morozov2014save}, there is a growing number of {\it hackathons for good} that explicitly invite participants to develop non-technological solutions \cite{taylor2018everybody}. Research on hackathons in general and hackathons for social good in particular has been growing in the past years. Most research on hackathons, however, has focused on in-depth studies of a limited number of events with the aim to making them more accessible beyond a tech-savvy audience \cite{irani2015hackathons}, fostering collaboration during an event \cite{trainer2016hackathon}, and increasing their impact, e.g., related to learning \cite{schulten2024we} and community building \cite{nolte2020support}. Recent research has shown how hackathons, often extending beyond technical matter, are increasingly considered as meaningful participatory processes to engage citizens \cite{dalpiaz_participation_2020}. Collaborative innovation, or co-innovation, is a social process of developing diverse ideas and the successful implementation of the best creative outputs \cite{George2007}. People who engage in co-innovation in hackathons are driven by various intrinsic motives, such as personal development (taking on a challenge, learning) akin to competence motivation in self-determination theory \cite{Rigby2018}, extrinsic benefits such as employment opportunities and prizes, as well as fun and enjoyment \cite{Rys2022}. Importantly, participants of hackathons for social good also report \textit{mission} or a contribution to a good cause as a primary motivation, suggesting prosocial motives.

\subsection{Neurobiological Basis for Purpose and Collective Innovation}

We propose that people volunteer to organise and participate in hackathons out of their own intrinsic motivation. The sense of contribution to a higher collective purpose, as well as various social and self-mastery related reasons to participate (such as the joy of meeting new people, opportunity to network and learn), are the primary forces driving the organisation of hackathons. Our thesis rests on the definition of intrinsic motivation as a natural drive for human action in theory of self-determination, according to which people are naturally drawn to work on activities that they find interesting, engaging and important. In other words, intrinsically motivated individuals spontaneously and voluntarily engage in work activities that are aligned with their inner motivational drive (such as curiosity or creativity \cite{Gagne2005}), as opposed to being driven by extrinsic goals, such as responding to the orders of superiors or the prospect of pecuniary benefits. 
In \cite{Lombard2024}, it was suggested that hackathons combine conditions in which individuals can generate deeply positive emotions, leading to a state of excitement, empowerment and satisfaction that makes them want to repeat the experience through a reward-learning mechanism. One can even attempt to characterise some of the neurobiological determinants of hackathons as kairos moments: (i) high social reward from meeting like-minded people who approve your ideas, a sense of autonomy, creative self-expression and play, which activate reward systems in the brain and release endogenous opioids \cite{Inagaki2018}, coupled with (ii) multiple sources of novelty which stimulate adrenaline release and uphold a state of optimal arousal and attention \cite{Aston-Jones2005}, while providing (iii) psychological safety and conditions to trust and cooperate, which releases oxytocin \cite{ Nave2015}, as well as (iv) optimal level of constraints that prevents excessive stress that could impair creativity \cite{Acar2019, Byron2010}.

A climate of psychological trust is necessary for participants to be willing to share their ideas and concerns, and to take risks in the innovation efforts \cite{Nembhard2006}. When people feel psychologically safe, they are more likely to trust each other. Cooperation critically depends on trust and humans are generally more cooperative than could be predicted by the rational economic theory \cite{Romano2017}. Indeed, collective action includes cooperative behaviours (where two or more people work toward a mutually beneficial outcome) and collective helping behaviours (where two or more people work for the benefit of others not involved in action) \cite{Ostrom2014}. 
In the physiologic model of Empathy-Collective Action, such action is motivated by empathetic concerns for others and necessitates the perception that others are in need \cite{Zak2013}. In this model, oxytocin induces empathetic concern that in turn increases the likelihood of collective action. Oxytocin is the hormone of trust, social memory and attachment \cite{Nave2015}. It is released during 
affiliative behaviours, such as enjoying the company of others, interacting with others and feeling part of a group, all of which have been reported by hackathon participants \cite{taylor2018everybody, Lombard2024, nolte2020support}. 
In addition, oxytocin reduces anxiety and promotes a ``tend and befriend'' response to stress, which, in contrast to confrontation \cite{Zheng2016}, is likely to appease social conflicts should they arise, and contribute to maintaining long-term cooperation.

Collaboration involves fluently attuned participants oriented toward a shared goal \cite{Mejia-Arauz2018}. Building consensus in a group depends on the motives for information sharing. When individuals desire to affiliate with others, they will likely seek consensus on personally important issues (such as the realisation of a sustainable development goal or climate action; \cite{Levine2018}). Emergent research on collaborative imagination shows that thinking about the future together facilitates the formation of new relationships because it leads to a deeper sense of closeness and a higher degree of social connection between strangers compared to performing a problem-solving task together \cite{Fowler2023}. This means that the task of co-innovation in a psychologically safe space reinforces the goal of building and maintaining social relationships. To maintain this focus on intrinsic motivation and cooperation, collaborative innovation events should avoid individual monetary incentives as they have been shown to have no effect on team collaboration in crowd-based innovation \cite{Riedl2017}. 

Finally, social diversity is conducive to creative thinking and innovation because newcomers in groups promote creativity in that they offer heterogeneous knowledge and perspectives \cite{Choi2005}. Paradoxically, the pressure to maintain a {\it status quo} and a professional identity is minimised in a team of socially diverse stakeholders because the utility of upholding related beliefs bears little external value \cite{Sharot2023}. Rather, the benefits of believing in the values shared with the group are now rewarded with social approval, which facilitates consensus building \cite{Georgiou2023}.\\

So far, insights into larger ecosystems of hackathons, and how series of hackathons contribute to community building and stewardship are scarce. In this paper, we address this gap by investigating a large number of {\it hackathons for good}, which can be related to SDG challenges associated with some of the pressing challenges faced by humankind, and which concern most of the actors of the multilateral system. We also consider how {\it hackathons for good} have contributed to the development of a new form of {\it computational diplomacy} based on the massive, participatory, decentralised production of open source software and beyond source code, possibly all sorts of other digital artefacts.

\section{Theory and Hypotheses}
\label{sec:theory}

We theorise that {\it Computational Diplomacy} is a decentralised participatory process, which involves collectively designing software for the sake of advancing the grand challenges of the world. One core engine of computational diplomacy is a succession of special \textit{kairos} events, which not only contribute to highly non-linear short-term production and long-tailed long-term production, but also play an important role in the recruitment of new community members. Hence, these special events punctuate the highly non-linear growth and production dynamics of computational diplomacy as a complex adaptive system, similar to instances of collective action in open source communities \cite{sornette_how_2014,maillart_aristotle_2019,muric_collaboration_2019}. 

\paragraph*{\textbf{Hypothesis 1 :} Productivity \& Engagement.}
We consider the \textit{productivity} of computational diplomacy events, compared to similar special events in terms of relatedness to grand challenges, their productivity in terms of software projects, and their capacity to gather crowds beyond their usual communities. For that, we propose two sub-hypotheses:

\begin{itemize}
    \item {\it H1a}: The number of computational diplomacy events has significantly increased over time.
    \item {\it H1b} : Computational diplomacy events attract a significant proportion of newcomers, and hence contribute to rapidly expanding decentralised participatory communities, while continuously providing perspectives to the collective action process.
\end{itemize}

\paragraph*{\textbf{Hypothesis 2 :} Critical Cascades in Computational Diplomacy.}

Investigating further the long-term effects of computational diplomacy events beyond their immediate kairos moments, we hypothesise that computational diplomacy is developing as a complex adaptive system operating at, or close to, criticality and subject to a succession of exogenous shocks \cite{crane_robust_2008}, followed by long-memory decays \cite{sornette_endogenous_2004}, which contribute to shaping the computational diplomacy ecosystem. The contribution of each special event as an exogenous shock writes,

\begin{equation}
  A(t) \sim (t-t_c)^\alpha,
  \label{eq:decay}
\end{equation}

with $t_c$ the critical time around peak activity, $\alpha = 1 \pm \theta$ and $\theta \approx 0.40$ being the renormalisation exponent \cite{sornette_critical_2006}. In case $\alpha = 1+\theta > 1$, the cascading process is sub-critical. Alternatively, if $\alpha = 1-\theta < 1$, special events trigger critical cascades \cite{crane_robust_2008} and thus, long-memory processes \cite{bouchaud_anomalous_1990}. 

\paragraph*{\textbf{Hypothesis 3 :} Productivity across special events.} 
We assume that computational diplomacy events are largely organised in a decentralised manner, and certainly with no central planner. Yet, we are still keen to understand how the events may be related to each other. For that, we test the statistical properties of participants and projects returning across computational diplomacy events. Similarly to the cascading process argument developed in \textit{Hypothesis 2}, we hypothesise that some participants exhibit large deviation cascades of engagement in multiple computational diplomacy events.
\section{Results}
\label{sec:results}

Considering a decentralised, bottom-up and scalable participatory process for the future of multilateralism, we show how a series of punctual {\it hackathons for good} have played an important role in building a community of participants engaged in computational diplomacy through the development of digital code bases dedicated to addressing one or multiple SDGs. We investigated (i) how hackathons are indeed directed towards tackling the SDGs, (ii) how hackathon participants make use of shared and differential technological solutions to contribute innovative advancements for the SDGs, (iii) how hackathons uniquely shape the dynamics of contributions on Github [Section \ref{sec:exo_critical}], and (iv) how hackathons largely contribute to on-boarding new participants [Section \ref{sec:new_participants}].

\subsection{Hackathons as temporary technological hubs for tackling global challenges}
\label{sec:sdg}

\begin{figure}
    \centering
    \includegraphics{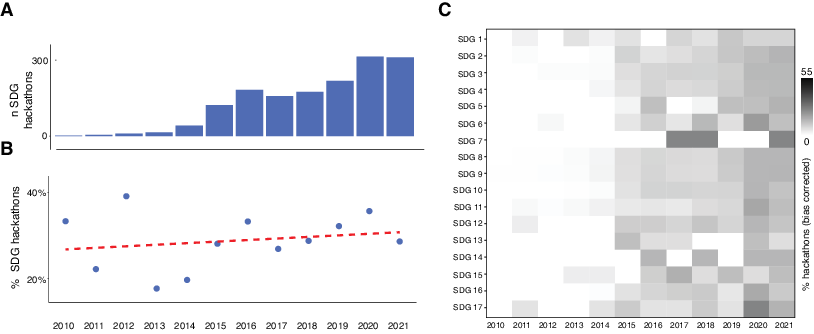}
    \caption{{\bf Hackathon alignment with Sustainable Development Goals.} {\bf A.} Number of SDG-related hackathons over time. {\bf B.} Percentage of SDG-related hackathons over time and linear fit trend(red dashed line). {\bf C.} Percentage split by SDG.}
    \label{fig:sdg}
\end{figure}

First, we ought to decipher which hackathons in our dataset are aligned with the SDGs. Applying an ensemble LLM method for mapping SDG contents in hackathon descriptions \cite{meier2022text2sdg}, we found that $\approx 30\%$ of them hold significant semantic relation with at least one of the 17 SDGs, proportion which is stable over time (Figure \ref{fig:sdg}{\bf B}). Indicative of this stability, a linear fit on the temporal progression in SDG-related hackathons is not significant but only indicative of a very slight increase over time (s=0.36, R2=0.04, p=0.53). Figure \ref{fig:sdg}{\bf A} gives further details on the evolution of the number of SDG-related hackathons, indicating trend increases in 2015 and 2020. Examining each SDG independently (Figure \ref{fig:sdg}{\bf C}), we observed that, across years, SDGs 4, 9, 8 and 10 are most covered in hackathons: quality education ($24.3\%$), industry, innovation and infrastructure ($21.6\%$), decent work and economic growth ($20.1\%$), and reduced inequalities ($12\%$) (temporal bulk not illustrated). Tellingly, one can note that SDG-related considerations have been embedded in the practice of hackathons since at least 2010, i.e., long prior to the launch of the SDGs by the United Nations in 2015. While most SDGs are represented in hackathons stably across years, some exhibit bursts of engagement, such as SDG 7 (Affordable and Clean Energy) in the years 2017, 2018 and 2021.

%\subsection{Shared and differential technologies for the common good}
%\label{sec:sdg_tech}

\begin{figure}
    \centering
    \includegraphics[width=.9\textwidth]{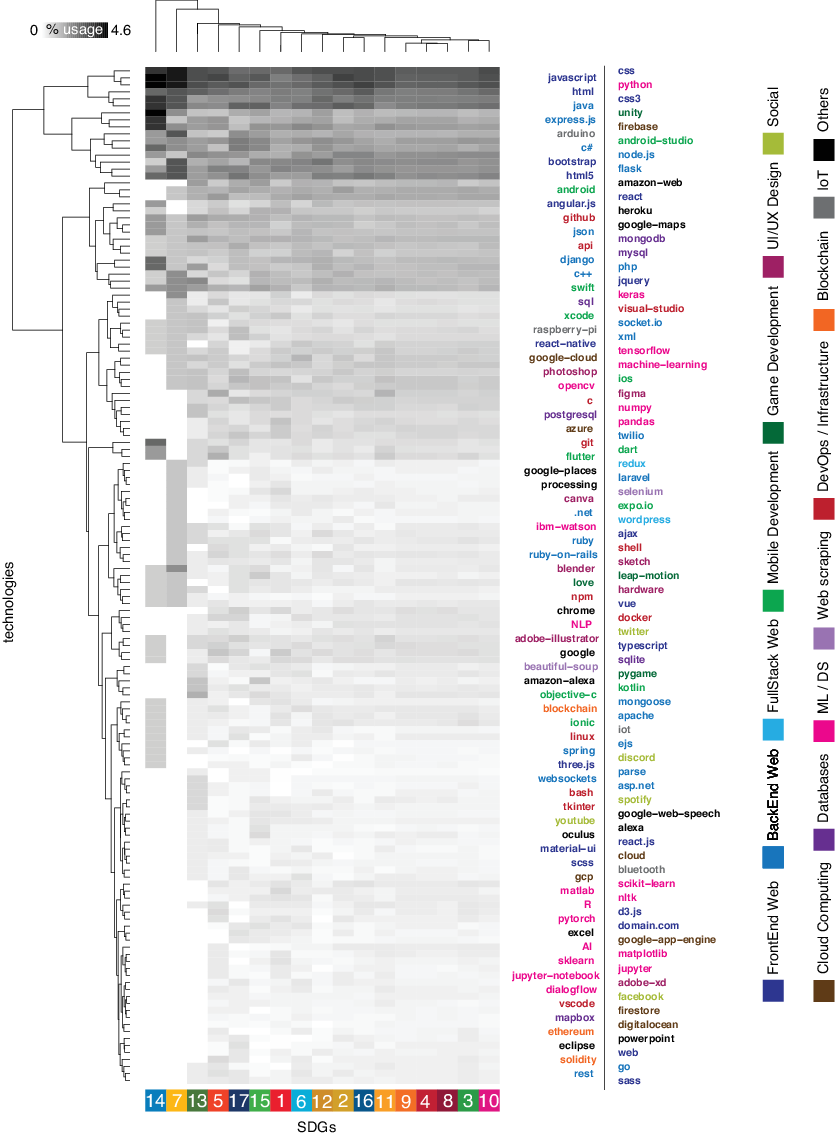}
    \caption{{\bf Distribution of hackathons using specific technology sets.} Heatmap gradient color-coding from 0\% (white) to 4.6\% (black). X-axis displays one column per SDG, y-axis shows one row per technology. Technologies are color-coded by major categories (see right legend).}
    \label{fig:sdg_tech}
\end{figure}

Next, we investigated whether the development of SDG-aligned software solutions through hackathons make use of specific or general purpose, thus shared, technologies. To do so, we used unsupervised machine learning (hierarchical clustering) to disentangle patterns in SDG-technology enrichments (see Data \& Methods in Supplementary materials). Results presented in Figure \ref{fig:sdg_tech} indicate that the same set of core general purpose technologies are the most permanently used for hackathons addressed to all SDGs: css, javascript, html, python and java, lead for all SDG-related hackathons (see top portion of Figure \ref{fig:sdg_tech}). Solutions developed for SDGs 7 (affordable and clean energy), 13 (climate action), and 14 (life below water) make large use of distinctive technologies (see left most portion of the heatmap in figure \ref{fig:sdg_tech}). For example, solutions for SDG 14 make an unusual high use of mobile development software tools, such as {\it flutter} and {\it dart}. Note that the technology usage per SDG is $<5\%$, showing a large diversity.

\subsection{Hackathons as exogenous critical shocks}
\label{sec:exo_critical}

\begin{figure}
    \centering
    \includegraphics[width=1.\textwidth]{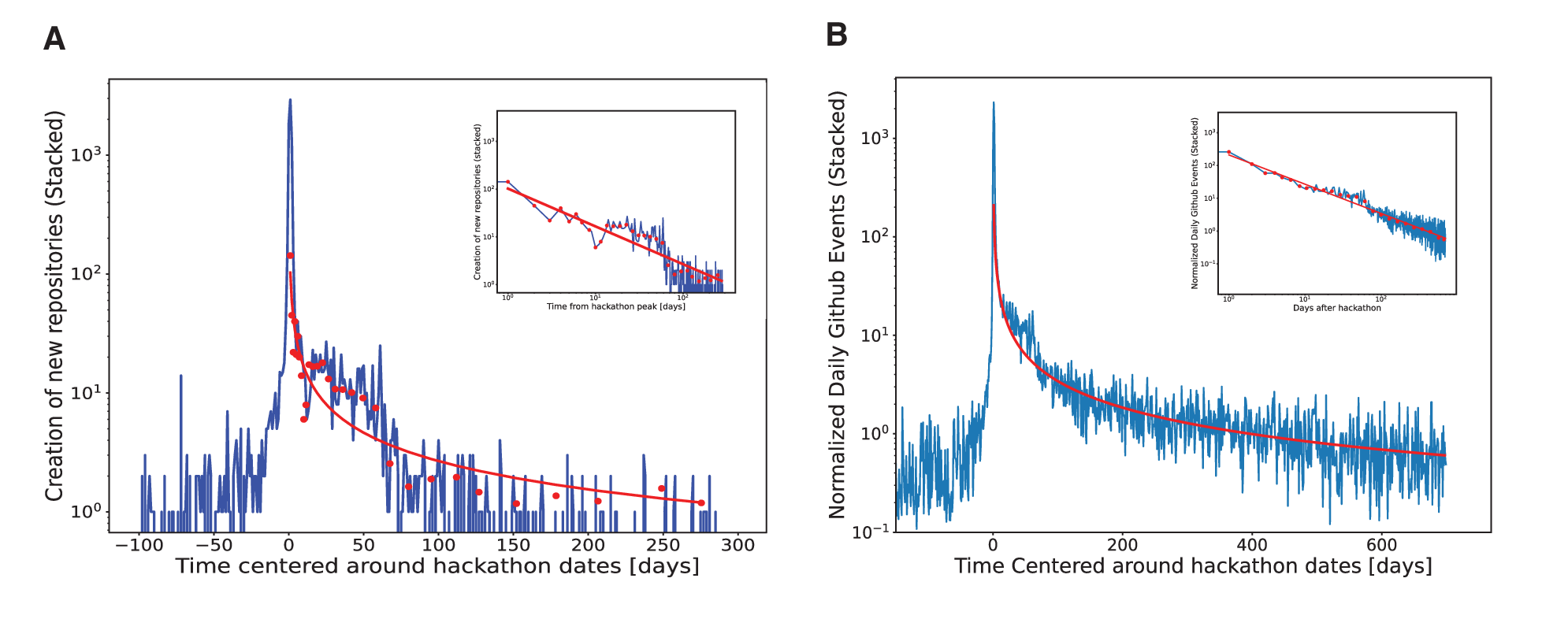}
    \caption{{\bf GitHub dynamics surrounding hackathon dates (stacked over all hackathons).} {\bf A.} Number of repositories created stacked over all hackathons. {\bf B.} Average events created during and after hackathons (event amplitude normalised). $t=0$ is the hackathon starting date.}
    \label{fig:hack_peak_decay}
\end{figure}

One central hypothesis of the computational diplomacy theory is that computational diplomacy events (e.g., hackathons) are catalysing moments of intense activity, but also with long-term implications. Figure \ref{fig:hack_peak_decay}{\bf A} shows the average number of repositories (repos) created around hackathon events (stacked over all events with origin $t=0$ set at the starting day of the hackathon). Many repositories are created during hackathons, followed by a slow power-law relaxation following equation \ref{eq:decay}, with decay exponent $\alpha_{repos} = 0.792(9)$ ($p < 0.001$ and $R=0.923$; c.f., inset). The value $\alpha_{repos} = 0.792(9)$ indicates that events tend to trigger exogenous cascades of repo creations that are close to criticality. Hence repo creation dynamics last in time following a hackathon, with new repos created up to hundreds of days after the event. Figure \ref{fig:hack_peak_decay}{\bf B} shows the average Github event activity performed on repos created during SDG hackathons (stacked over all hackathons with peak activity within a 3-day range of the hackathon date; excluding repos created 3 days after the starting date of the event). Although event activity is maximum during the hackathon, it is on average declining slowly over at least 700 days, following equation \ref{eq:decay} with $\alpha_{events} = 0.876(3)$ ($p < 0.001$ and $R=0.995$; c.f., inset). Although it is still a very slow decay response with $\alpha_{events} < 1$, this exponent is slightly higher compared to what would have been expected for exogenous critical cascades ($\alpha = 1 - \theta$ with $\theta \approx 0.4$) \cite{sornette_endogenous_2004}. This may be a consequence of some mixing between exogenous critical and sub-critical cascades. Disentangling the two regimes would require further investigation following, e.g., \cite{crane_robust_2008}. Nevertheless, the slow decay ($\alpha < 1$) indicates that beyond the jolt of kairos moments, hackathons have on average long lasting impact on the development of software and hardware projects, possibly for over several years. 

\subsection{New participants in hackathons}
\label{sec:new_participants}

\begin{figure}[hb!]
    \centering
    \includegraphics[width=1.1\textwidth]{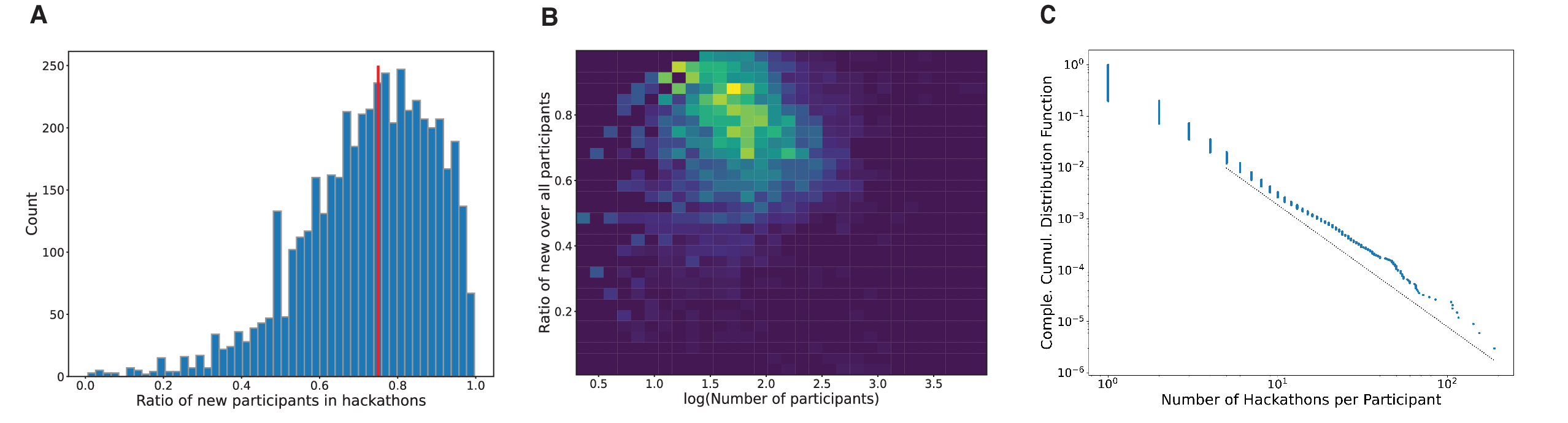}
    \caption{{\bf Hackathons as attractors of new participants and recurrence.} {\bf A.} Count of hackathons by ratio of new participants. {\bf B.} Ratio of new participants as a function of hackathon size ($n$ participants). {\bf C.} Complementary cumulative distribution function of the number of hackathons attended per participant.}
    \label{fig:renewal_and_retention}
\end{figure}

One key requirement of the shifts advocated for the multilateral United Nations system is the inclusion of a large number of citizens, possibly all Earth citizens. We therefore studied the proportion of new participants per hackathon as a gauge of reach potential. Figure \ref{fig:renewal_and_retention}{\bf A} shows the count of hackathons with a given ratio of new participants. The median percentage of participants that have not participated in a SDG hackathon beforehand is $72.6\%$. This shows that hackathon events are strong attractors to new participants. Figure \ref{fig:renewal_and_retention}{\bf B} shows the ratio of new participants as a function of hackathon size in terms of participants. One sees that the ratio of new participants is maximum ($\approx 0.595$) for hackathons of around $25$ participants and decreases to $\approx 0.7$ for hackathons of size $100$. Figure \ref{fig:renewal_and_retention}{\bf C} shows the complementary cumulative distribution function (CCDF) of the number of hackathons attended per participant. The CCDF exhibits a power-law 

\begin{equation}
    P(X > x) \sim 1/t^{\mu},
    \label{eq:powerlaw}
\end{equation}

with exponent $\mu_{participants} = 2.37(2)$ ($x_{min} = 4$, $max_{llh} = 5.63\cdot 10^{5}$). Because $\mu_{participants} > 2$, its two first statistical moments are stable \cite{sornette_critical_2006}. Therefore, the distribution is only slightly fat-tail. It nevertheless shows that a majority of participants only attend to one or few hackathons although some participants return regularly to hackathons, with a few participants having attended more than $100$ hackathons.

\section{Discussion}
\label{sec:discussion}
Using a unique combination of two datasets (i.e., Devpost and Github Archive), we have performed a partial test of the computational diplomacy theory. We investigated how $1,320$ hackathons with mention of SDG related keywords (out of $4,885$ hackathons have fuelled the development of software code that contributes to address the sustainable development goals (SDGs) identified by the multilateral system. Over the study period from 2010 until 2022, we found that although the number of hackathons has linearly yet tremendously increased by $3\%$ per week, the proportion of hackathons with SDG concerns have remained stable around $30\%$ of all hackathons. Over time and particularly since 2020, nearly all SDGs have been increasingly consistently addressed in the development of software code, at the notable exception of SDG 7. Fifteen technology categories and tens of software and hardware technologies have been used. We also found that computational diplomacy hackathon events not only play a significant role in fuelling activity in terms of code repository creation and activity on these repositories on the short-term and on the long-term, they also contribute to the massive recruitment of new participants although the measured retention rate of participants does not suggest critical cascades of participants to multiple hackathons. Although we acknowledge that by sheer sampling effect our study probably largely underestimates the size of the community engagement in writing code for the SDGs, it nevertheless makes the case for a largely under the radar build-up of computational diplomacy. This is similar to the open source software movement, which remained a niche \cite{levy_hackers_2010} until it became mainstream and its spirit has percolated in society \cite{benkler_penguin_2011}. Since SDG-related hackathons started to occur before the SDGs, one may further envision the computational diplomacy community as a precursor of a much larger movement in tackling grand challenges. This is a powerful message of hope that decentralised peer-production can actually identify and undertake action at scale somehow {\it before} the institutions.

While computational diplomacy has shaped the future of the multilateral system through the production of millions of lines of code, quantity matters far less than its qualitative implications for radical change. We suggest that hackathons trigger collaborative or collective kairos moments, i.e. shared experiences of a significant and opportune moment in time that can lead to a collective awakening, realisation, or a transformative event. Here, we take a psycho-biological perspective to advance some hints on why computational diplomacy as an ecosystemic participatory process \cite{mahajan_participatory_2022} can work through the decentralised organisation of hackathon events. Further, to underscore the importance of computational diplomacy, we outline three concrete cases of deep relevance for multilateralism.

\subsection{Why computational diplomacy can work? A psycho-biological perspective}
\label{sec:psychobiological_perspective}

%{\bf [we need to cite and expand on our play and work paper \cite{Lombard2024}, posted on SSRN (\url{https://papers.ssrn.com/sol3/papers.cfm?abstract_id=4783197})]}

Although today's grand challenges imply the development of technologies, economies and social practices to overcome deadlines that are somewhat existential to humanity, they remain hard ``wicked'' problems that require much collective intelligence and a step-by-step modular bootstrapping approach \cite{henderson_innovation_2021}. But they foremost involve aligning the motivations of different actors in order to mobilise cross-collaborative innovation. As we showed in this work, distributed modular innovation events, such as hackathons, have the potential to serve as starting points for such innovations. Research shows that hackathons present the right conditions to align motivations, create a shared reality, and build trust and social contracts between participants to ultimately ensure long-term cooperation \cite{Lombard2024,nolte2020happens}. To best capitalise on the desire to contribute to a social cause, participants must maintain intrinsic motivation \cite{Gagne2005} and have the autonomy to align their work with the higher purpose they aspire to, such as contributing to the greater good \cite{DiBenigno2022}. Because such participation provides meaning, people who work with such motivation tend to be highly satisfied, have high levels of performance \cite{Rigby2018}, and hence, are more likely to act pro-socially \cite{Grant2008}. People acting with prosocial motivation produce more useful (socially desirable and needed) innovations \cite{Grant2011}.

An effective way to combine and sustain intrinsic motivation, a sense of autonomy and trust among diverse participants, is to allow and encourage them to \textit{play}. Indeed, the motivation to engage in play-related behaviours such as experimenting in a safe space and interacting with others in unusual ways has been reported as the main driver for most social good hackathon participants \cite{Lombard2024}. Co-innovation events should encourage social play – playful activities in which the target of the interaction is other individuals \cite{Petelczyc2018} because this kind of play helps boost innovation \cite{Mukerjee2022}, maintains group cohesion and encourages cognitive flexibility in adults \cite{Pellis2013}. The performance of social play is modulated by the neurotransmitter systems intimately implicated in the motivational, pleasurable and cognitive aspects of natural and drug rewards, such as opioids, endocannabinoids, dopamine and norepinephrine \cite{Trezza2010}. Therefore, when people play, they are more likely to associate the experience with rewards that will make them want to continue and return to it, as well as to attain mood states that are conducive to creativity, such as positive emotional states, especially when coupled with an activating emotion, which reduce switch costs while enhancing the performance in divergent thinking and problem solving \cite{Khalil2019}. Furthermore, co-creating in a playful way reinforces the positive experiences and all the psychological states mentioned above (self-validation, autonomy and intrinsic motivation, affection for team members, shared ownership over the developing idea \cite{Rouse2020}).

Future research on the importance of hackathons in computational diplomacy should further explore the link between participant motivation, play, levels of trust and collective action outcomes for social change. It is also worthwhile to explore how such co-innovation events foster collective imagination and create a common vision of a desirable future\cite{HAWLINA202031}. 

\subsection{Three concrete cases for computational diplomacy}
\label{sec:concrete_cases}

Our results show the power and possibly the importance of ecosystemic participatory decentralised processes driven by special kairos moments of collective intelligence for purpose, with possible psycho-biological ramifications. Here, we consider three concrete computational diplomacy cases for the future of multilateralism,ranging from the production of software code and hardware code to building and maintaining peace, to radically transforming multilateralism to resilient to a digitised and fast changing world.

\subsubsection{Peer production of software code and hardware designs}

Shift 2 of the UN Summit of the Future focuses on supporting a just digital transition that unlocks data value and protects against digital harms. Participatory decentralised processes are vital for fostering collaboration, innovation, and problem-solving, bringing together diverse stakeholders to tackle common challenges. These processes promote cross-fertilisation of ideas and innovative solutions in a transparent and accountable environment , which is essential for a sustainable and equitable world.

Hackathons are effective at integrating diverse perspectives into actionable knowledge by allowing people to collaboratively explore and develop new ideas. For example, a 2023 hackathon on water quality monitoring, organized by the World Meteorological Organisation and other international bodies, brought together over 60 experts globally to generate novel solutions \cite{chernov_innovative_2024,warner_empowering_2024,lopez-maldonado_contributions_2024, cacciatori_gems_2024}. Participants praised the hackathon format for its ability to explore radical ideas and create prototypes efficiently, surpassing traditional bureaucratic methods. And as a token of efficiency and ease of replication, some participants already organised their own local hackathons (e.g., in Brazil). The computational diplomacy hackathon approach aligns with multilateralism by fostering shared responsibility and purpose, accelerating creative solutions, and promoting collective intelligence and innovation.

By remaining decentralised and fostering community engagement and shared practice \cite{von_krogh_carrots_2012}, hackathons offer a fertile ground for a transparent and accountable form of multilateralism through kairos temporary organisation \cite{orlikowski_its_2002}, hence avoiding disruption of legacy institutions. They leverage individual expertise and cultural diversity while focusing on collective goals for the common good. In doing so, hackathons support a digital and data governance shift and enhance legitimacy and effectiveness through inclusion and accountability.

\subsubsection{Conflict Resolution} 

We then consider {\it Shift 5}, focusing on {\it Peace and Prevention: empowering effective, equitable collective security arrangements}. In conflict prevention and resolution, contact theory \cite{allport_nature_1954} offers a strong approach to reducing conflict between individuals and communities by fostering positive interpersonal interactions under specific conditions: equal status, shared goals, inter-group cooperation, and authoritative support. Originally developed to address racial and ethnic tensions, contact theory is now used in conflict resolution to create structured interactions that promote understanding and build positive relationships, laying a foundation for lasting peace \cite{mckeown_intergroup_2017}. Hackathons align well with contact theory by offering structured environments where diverse groups collaborate on shared goals, promoting mutual understanding and reducing prejudice. By bringing together people from different backgrounds on equal footing and valuing each participant’s contributions, hackathons foster equity and mutual respect. The collaborative nature encourages teamwork, breaks down barriers, and builds positive relationships. Thus, hackathons for good embody contact theory in action, facilitating meaningful interactions that reduce biases and promote social cohesion. In an era of growing internal and external societal challenges, such as polarisation and climate-induced migration, there is a pressing need to develop methods for conflict prevention and resolution. Computational diplomacy, which leverages collective intelligence, empathy, diverse perspectives, and shared purpose, extends beyond open-source collaboration to reduce interpersonal prejudice, a significant barrier to consensus and social justice.

\subsubsection{Transformational Change} 

As we, humanity, undertake an unprecedented experiment in altering our world across environmental, technological, and social dimensions, we must view it through the lens of its strongest drivers of change. Digitisation offers significant leverage for quantitative change, but moving from adaptive to transformative capacity requires proactive adaptation. Transformative change involves a fundamental shift in the structure, culture and operations of organisations, by rethinking processes to improve performance and sustainability \cite{novalia_theorising_2020,hawkins_leadership_2021}. The six fundamental shifts proposed for the UN Summit of the Future necessitate transformative adaptation to build equitable and sustainable societies through intentional planning and governance. International organisations need to go beyond traditional policy cycles, focusing on vision, planning, and governance, for which they were not originally designed \cite{newman_multilateralism_2009}. Transformative adaptation, whether to climate change or digital transformation, exemplifies challenges requiring global-scale responses. It demands a shift from viewing problems as obstacles to seeing them as dynamic relationships, involving both being and doing. This approach seeks to establish new norms and frameworks while fostering an eco-systemic awareness over individualism.

The lack of a shared purpose in our civilisation means we struggle to implement key performance indicators from agreements like the Paris Agreement or the SDGs due to divisive interests. Current collective intelligence in international relations is limited, as achieving objectives often requires a change in behaviours, thinking, and partnerships beyond existing capacities \cite{hawkins_leadership_2021}. Hackathons already draw thousands globally to address SDG-related topics, presenting a choice for the multilateral system: dismiss or embrace computational diplomacy. Dismissing it risks losing innovative solutions and motivated individuals; embracing it means integrating its outcomes at national, regional, and global levels. Hackathons, like participatory design, have the potential to co-design the multilateral system, hence profoundly altering its fundamental mechanisms \cite{bodker2004participatory, fischer2004meta}. To avoid disruption, an initial step could be incorporating hackathons into UN processes and supporting promising projects within or outside the UN framework.

\section{Conclusion}
\label{sec:conclusion}

In this article, we theorised computational diplomacy as a decentralised participatory process where individuals collaborate to design software and hardware aimed at tackling the world's most pressing challenges. This approach relies on a series of kairos events —critical moments that drive both intense bursts of short-term innovation and sustained, long-term contributions. These events are vital for fostering new ideas and recruiting participants, highlighting the non-linear growth and adaptability of computational diplomacy, similar to patterns observed in open-source communities. Hackathons exemplify these kairos moments, bringing together diverse groups to address UN SDG topics through various technologies. Our research shows that 30\% of hackathons since 2010 have focused on these goals, driving the long-term creation of new projects and developing community engagement. Beyond immediate outcomes, computational diplomacy hackathons leverage the neurobiological basis of human cooperation and empathy, fostering a collective sense of purpose. They create inclusive spaces that encourage collaboration and innovation, contributing to a more participatory model of digital governance. This bottom-up approach enhances diversity, reduces prejudice, and fosters solutions through continuous peer engagement and review.

This study of computational diplomacy is a first of its kind, combining a new theory and partial empirical evidence. It takes a radical perspective on multilateralism. Yet this perspective is strongly aligned with the fundamental digitisation mechanisms of society that have occurred in the last five decades. As such it bears limitations on data (depth of exploitation of the current dataset and sample size, which could be increased in the future). Besides data, we contemplate two major limitations, which shall be also be seen as venues for future research: (i) the nature of produced digital artefacts, their iterative improvement through peer-production, and their real impact on contributing to solving grand challenges remain largely unclear; and (ii) the neurobiological basis for purpose and collective innovation would deserve deeper scrutiny. Advancing research at the interface of neurosciences, psychology and collective intelligence is an outstanding scientific challenge. Better characterising these mechanisms could help design physical and digital experiences to overcome some of the barriers impeding effective responses to the grand challenges faced by humankind. 

To conclude, computational diplomacy, grounded in its neurobiological basis and driven by kairos moments, offers a transformative and inclusive model for digital governance. By harnessing the collective intelligence of a global community, it presents a resilient framework for addressing the complex challenges facing our world today and offers a credible perspective for envisioning the multilateral system of the future.


\textbf{Acknowledgements: }The development of this manuscript would not have been possible without the valuable input from interactions with attendants to the conference on {\it Co-Creating the Future: Participatory Cities and Digital Governance} held on September 11-13, 2023 at the Complexity Hub (Institute of Advanced Studies) in Vienna, Austria. This article was further inspired by a one-day hackathon, which was held at the {\it UN Library \& Archives Geneva} in Geneva, Switzerland on November 8, 2023 as part of the week-long Hack the Hackathon workshop\footnote{\href{http://hackthehackathon.org/}{http://hackthehackathon.org/}} which was partially funded by the German Federal Government Commissioner for Culture and Media as part of the CIRCE initiative\footnote{\href{https://creativeimpact.eu/en/}{https://creativeimpact.eu/en/}}. The aim of the one-day hackathon was to explore how hackathons could offer opportunities for the development of radical forms of multilateralism.

\bibliographystyle{unsrt}  
\bibliography{bibliography.bib}  %%% Remove comment to use the external .bib file (using bibtex).
%%% and comment out the ``thebibliography'' section.

\end{document}